# Astro2020 Science White Paper

# The production and escape of ionizing photons from galaxies over cosmic time

**Thematic Areas:** ☐ Planetary Systems  ☐ Star and Planet Formation
☐ Formation and Evolution of Compact Objects  ☐ Cosmology and Fundamental Physics
☒ Stars and Stellar Evolution  ☐ Resolved Stellar Populations and their Environments
☒ Galaxy Evolution  ☐ Multi-Messenger Astronomy and Astrophysics

**Principal Author:**
Name:         Dr. Jane Rigby
Institution:  NASA Goddard Space Flight Center
Email:        Jane.Rigby@nasa.gov
Phone:        301-286-1507 (office)

**Co-authors:**
Danielle Berg (OSU)
Rongmon Bordoloi (NCSU)
John Chisholm (UCSC)
Michael Florian (NASA GSFC, USRA)
Matthew Hayes (Stockholm U.)
Michael Gladders (U. Chicago)
Bethan James (STScI)
Sangeeta Malhotra (GSFC)
Sally Oey (U. Michigan)
John M. O'Meara (W. M. Keck Observatory)
T. Emil Rivera-Thorsen (ITA-UiO)
Keren Sharon (U. Michigan)

**Abstract:**
The ionizing photons produced by massive stars are key actors in galaxy evolution.  Ionizing photon production and escape is poorly understood.  Improved space-based, spatially-resolved, multiplexed spectroscopic capabilities covering $\lambda_{obs}$ = 1000-3000 Å, in concert with spectroscopy from the ELTs and *JWST*, would lead to definitive answers as to how ionizing photons are produced and leaked, what populations of galaxies are responsible for ionizing photon leakage, what determines whether escape is possible, and how ionizing galaxy populations evolve over cosmic time.

**Key Science Questions**:
- How do galaxies produce ionizing photons?
- How do ionizing photons escape from galaxies?
- Which galaxies leak ionizing photons? Which galaxies reionized the Universe?
- What about those galaxies permitted ionizing photons to escape?

**Ionizing photons are important.**

The Universe was re-ionized; therefore there must have been enough ionizing photons to do the job, and to keep the Universe ionized at later epochs. Studying galaxies in the epoch of reionization is a key science goal for JWST, which will characterize the galaxy luminosity function and measure physical properties like densities and metallicities. However, JWST will be unable to measure the ionizing fluxes of galaxies at high redshift, for a simple reason: the opacity of the intergalactic medium prevents ionizing photons from $z \lesssim 4$ from reaching Earth. Only at redshifts $z \lesssim 4$ is it possible to measure ionizing flux from galaxies (see Fig. 9 of Steidel et al. 2018).

Ionizing photons are produced by main sequence and evolved massive stars. The ionizing photons ionize the H II regions, power the strong nebular emission lines that are key diagnostics of galaxy evolution, and if there are clear channels, escape their parent galaxy. The massive stars that produce the ionizing photons also produce the non-ionizing ultraviolet continuum, nucleosynthesize the heavier elements, and distribute those elements through a combination of stellar winds and supernovae winds.

After heroic effort, 36 galaxies have been spectroscopically identified to "leak" ionizing photons, spanning redshifts from z~0 to z~3 (Leitet et al. 2011; Borthakur et al. 2014; Vanzella et al. 2015, 2016, 2018; Izotov et al. 2016 a,b; Leitherer et al. 2016; Shapley et al. 2016; Bian et al. 2017; Izotov et al. 2018 a, b; Steidel et al. 2018; Rivera-Thorsen et al. 2019; **Figure 1**). We are just beginning to learn what kinds of galaxies might have reionized the Universe.

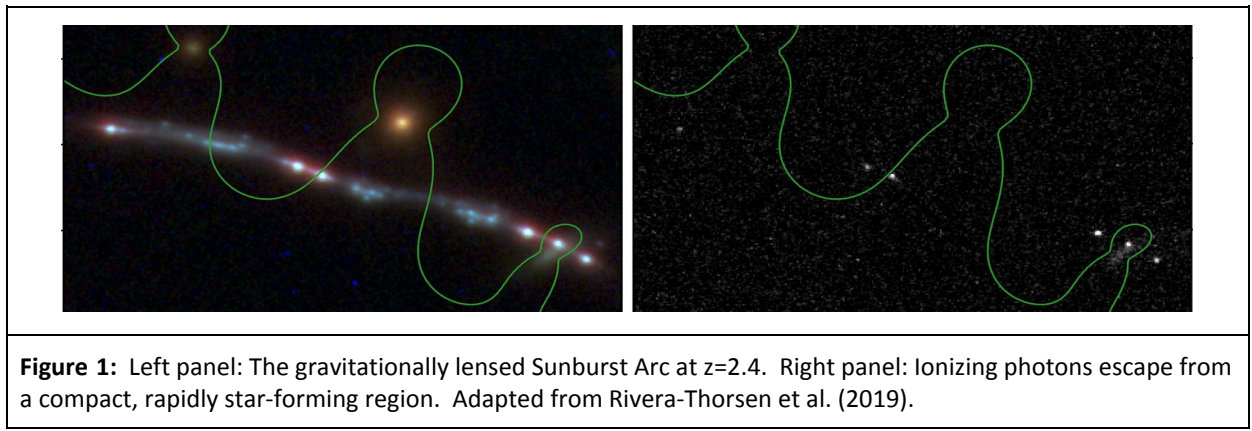

**Figure 1:** Left panel: The gravitationally lensed Sunburst Arc at z=2.4. Right panel: Ionizing photons escape from a compact, rapidly star-forming region. Adapted from Rivera-Thorsen et al. (2019).

**Massive stars make ionizing photons.**

Knowledge of the non-ionizing UV output of galaxies comes from either resolving individual massive stars in nearby galaxies (e.g. Crowther et al 2016), or by stellar population synthesis fitting to the integrated non-ionizing UV spectrum of a galaxy.  The non-ionizing UV spectra of galaxies features a rich suite of diagnostics of the hot stars, namely broad He II emission, the P Cygni profiles of C IV, Si IV, and N V, and photospheric absorption lines (deMello et al. 2000; Crowther et al. 2016).  These diagnostic features can be compared to stellar population synthesis (SPS) codes (Leitherer et al. 1999, 2014; Eldridge et al 2009, 2017; Conroy et al. 2009; Conroy & Gunn 2010), to determine the age and metallicity of the massive star population; at present, this can be done at z~2–3 only for galaxies that are gravitationally lensed (Pettini et al. 2000; Eldridge & Stanway 2012; Chisholm et al. 2019), or by stacking the spectra of many galaxies (Steidel et al. 2016.)  While SPS codes have made progress over the past two decades, they reflect key lingering unknowns about massive stellar evolution, including the role of binary evolution pathways (Stanway et al. 2016), the amount of stellar rotation (Levesque et al. 2012), the upper end of the IMF, and the winds, spectra, and fates of stars with very low metallicity (Ramachandran et al. 2019).  One way to improve these models is to obtain spectra of very low metallicity massive stars, which should happen through the STScI Director's Hubble UV Legacy Library of Young Stars (ULLYSES).  A second way would be to improve the theoretical spectra of hot stars that are derived from stellar atmosphere models, by increasing the spectral resolution to R>5000 to match in-hand data, and by adding non-solar alpha-to-iron abundance patterns.

Different stellar population synthesis models, when fitting the same high-quality spectra, differ by factors of several in the predicted ionizing photon production rate and the ionizing spectral slope $\lambda_{rest}$<900Å (Chisholm et al. 2019; **Figure 2**). In other words, the current unknowns about massive stellar populations translate to factors-of-several uncertainty in the expected ionizing photon production rates of galaxies.  We need to better characterize the massive stellar populations within star-forming galaxies, their intrinsic ionizing production rate, and what determines whether those ionizing photons are trapped or can escape.

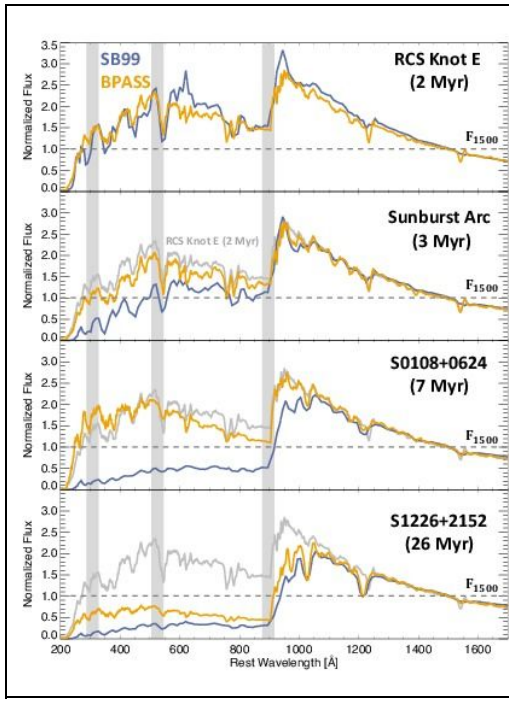

**Figure 2:** $\lambda_{rest}$= 1000–2000Å galaxy spectra reveal the age and metallicity of the massive stellar population, by fitting stellar population synthesis (SPS) models.  Current SPS models disagree by factors-of-several in the amount of ionizing radiation that should be produced by a given young stellar population.  This disagreement is due to fundamental uncertainties about massive stellar evolution and spectra. From Chisholm et al. (2019), using z~2 lensed galaxy spectra from Megasaura (Rigby et al. 2018a).

## Ionizing photons can escape galaxies. How?

Neutral H atoms absorb ionizing photons, which blocks their escape. To escape, ionizing photons can travel through clear channels (or highly ionized channels) in the gas, or can stream out of density-bounded H II regions that are smaller than their Strömgren radii (**Figure 3**). Ionizing photons from galaxy outskirts should have an easier time escaping (Keenan et al. 2017), and could play an outsized role.

It is not understood whether escape is determined by local factors (star formation history, nebular geometry and ionization) or large-scale factors (halo mass, galaxy mergers, environment), nor is the role of feedback understood. Do the massive stars break open their own neutral gas shells, like hatchlings, or is the neutral gas torn open by large-scale galactic events? How important is the higher UV background at higher redshift at weakening the shells? Do numerous small "eggs" dominate ionizing photon escape, or the most luminous ones?

The Lyman alpha profile is an indirect probe of the column density, geometry, and kinematics of the neutral gas (Verhamme et al. 2006; Jaskot & Oey 2014; Rivera-Thorsen et al. 2017b, Yang et al. 2016, 2017). Local strong leakers are dominated by narrow double-peaked profiles, indicating high ionization and low but non-zero H I column density (Izotov et al. 2018; Verhamme et al. 2015).

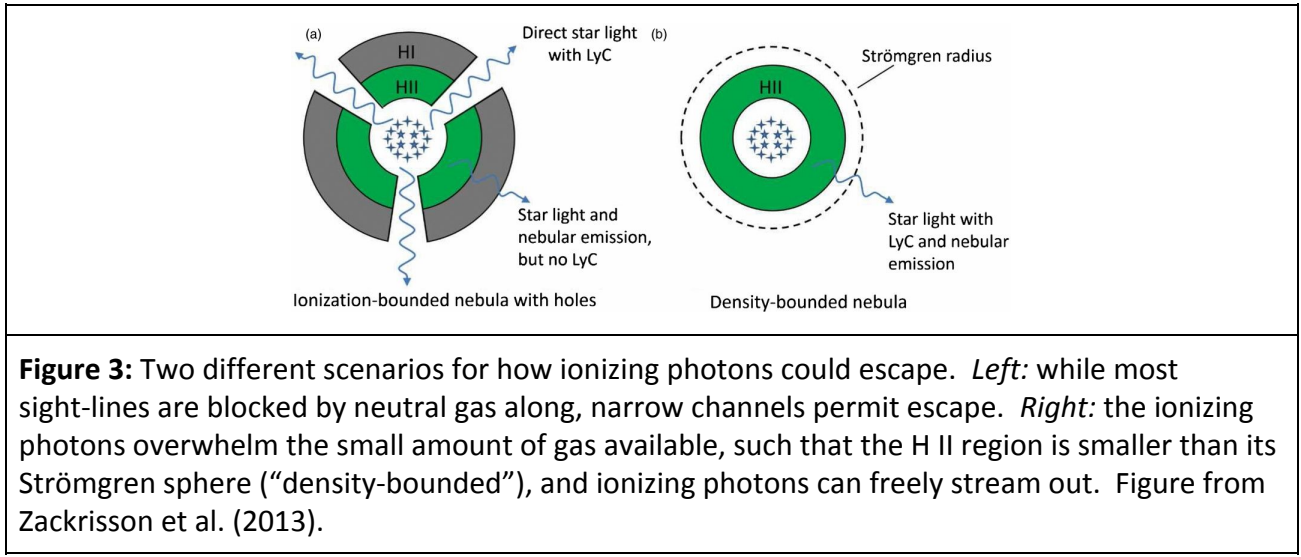

**Figure 3:** Two different scenarios for how ionizing photons could escape. *Left:* while most sight-lines are blocked by neutral gas along, narrow channels permit escape. *Right:* the ionizing photons overwhelm the small amount of gas available, such that the H II region is smaller than its Strömgren sphere ("density-bounded"), and ionizing photons can freely stream out. Figure from Zackrisson et al. (2013).

## Indirect predictors of escape:

How can we determine which high-redshift galaxies reionized the Universe, when the opacity of the intergalactic medium at high redshift prevents their ionizing photons from reaching Earth? Indirect indicators will need to be used, for example rest-frame optical emission line ratios (Jaskot & Oey 2013; Izotov et al. 2018), UV metal absorption lines (Rivera-Thorsen et al. 2015, 2017a; Gazagnes et al. 2018; Chisholm et al. 2018), Ly $\alpha$ (Verhamme 2015, Yang et al. 2017), and resonant emission lines such as C IV and Mg II (Henry et al. 2018; Berg et al. 2019). Thus, we must understand the processes of ionizing photon production and escape at z<3 where ionizing photons can be detected, and then transfer that understanding to spectra from JWST and the ELTs of galaxies in the epoch of reionization that capture spectral predictors of ionizing photon escape.

## How to answer the key science questions:

It is possible to definitively solve these key science questions, through a systematic investigation over cosmic time. It would then be possible to confidently extrapolate to the epoch of reionization, and use the measured properties of high-redshift galaxies to predict which ones are responsible for the bulk of the ionizing photons. This enterprise can be broken into four parts, divided by redshift:

**1) Redshift 0.1<z<1.5:** In this redshift range, covering most of cosmic time, both the rest-frame ionizing UV and the rest-frame non-ionizing UV can only be captured from space. The crucial measurements are sub-arcsecond spatially resolved spectroscopy that covers $\lambda_{rest}$ = 1000–3000Å at moderate spectral resolution (R=5000) and signal-to-noise (SNR $\geq$ 15 per resolution element) (Rigby et al. 2018a), and that covers the ionizing $\lambda_{rest}$ = 900Å region at much lower spectral resolution, R~100s. Critical goals that can only be done in this regime are to spatially resolve which star-forming regions are responsible for the ionizing photons, and measure the properties of each star-forming region (metallicity, stellar age, luminosity). Spatially resolving ionizing emission is the only way to directly determine the physical processes of escape, and to understand the interplay of mechanical and radiative feedback. Mapping Ly α emission (e.g., Ostlin et al. 2009) is also essential, as Ly α is a strong predictor of escape (Verhamme et al. 2015). It also makes sense to stretch the wavelength coverage as blue as possible, to measure the ionizing spectral slope at $\lambda_{rest}$<900Å, and thus constrain key unknowns about massive star evolution.

**2) Redshift 1.5<z<2.5:** Multi-object spectroscopic surveys with 20–30m ground-based extremely large telescopes (ELTs) will obtain the rest-frame non-ionizing UV spectra of large numbers of galaxies, at seeing-limited resolution, and thus characterize their massive stellar populations, as well as diagnostics of nebular conditions and gas inflow and outflows (e.g. Shapley et al. 2003; Pettini et al. 2000; Rigby et al. 2019b). The ELTS will also cover Ly α over most of this redshift range. However, the ELTs cannot measure the ionizing continua for any of these galaxies — this must be done from space, and would require sensitive, low spectral resolution, spatially resolved spectroscopy of these galaxies with a space telescope at $\lambda_{rest}$ = 900Å, which at these redshifts is $\lambda_{obs}$ = 2200–3150Å.

**3) Redshift 2.5<z<3:** The ELTs will capture the ionizing flux of large numbers of galaxies in the very blue optical ($\lambda_{obs}$ = 3200–3600Å), and their non-ionizing rest-frame UV spectra ($\lambda_{rest}$ = 1200–3000Å) in the optical and near-IR ($\lambda_{obs}$ = 4200–12000Å). This measurement would prove especially difficult if faint galaxies are responsible for the bulk of ionizing photon production. These measurements would be seeing-limited, and therefore would lack spatial information as to which parts of galaxies leak ionizing photons. Stacking Keck spectra of z~3 galaxies (Steidel et al. 2018) is a proof-of-concept that demonstrates how the ELTs can determine which galaxies are responsible for ionizing photon production and leakage in this redshift window.

**4) Redshifts 7<z<9:** In long integrations, the ELTs and JWST will obtain non-ionizing rest-frame

UV and rest-frame optical spectral diagnostics of galaxies at the epoch of reionization, including indirect predictors of ionizing photon escape. Subsequently, the investigations discussed in the previous 3 points would provide a much deeper understanding of ionizing photon production and escape for redshifts 0<z<3. It would then be possible to apply this knowledge to the ELT and JWST observations of the epoch of reionization, to comprehensively determine which galaxies reionized the Universe.

**To summarize,** the production of ionizing photons by massive stars, and their escape from galaxies, are key unsolved issues in our understanding of galaxy evolution and how the Universe was reionized. Only a systematic approach, using JWST and the ELTs at the highest redshifts, using the ELTs at 1.5<z<3, and a future observatory with spatially resolved space-based 1000-3000Å spectroscopic capabilities, can comprehensively measure the massive stellar populations within star-forming galaxies, determine which kinds of galaxies leak ionizing photons and how to predict that escape, and apply that knowledge to finally understand which galaxies ended the dark ages by reionizing the Universe.

| Facilities needed | Capabilities needed |
|---|---|
| *JWST* | Deep NIRSpec and MIRI spectra of rest-frame optical diagnostics of galaxies at reionization |
| ELTs | Multiplexed spectroscopic surveys covering $\lambda_{obs}$ = 3200-12000Å |
| Large space UV telescope (LUVOIR, HabEx) | Spatially-resolved spectroscopy that covers the ionizing $\lambda_{rest}$ = 900Å, $\lambda_{obs}$ = 1000–3000Å, at low spectral resolution (R~100s) |
| Large space UV telescope (LUVOIR, HabEx) | Spatially-resolved spectroscopy that covers $\lambda_{obs}$ = 1000–3000Å at R~5000 and SNR $\geq$ 20 per resoln. element |